\documentstyle[preprint,aps]{revtex}
\begin{document}
\draft
\title{Markov chain analysis of random walks on disordered medium}
\author{Sonali Mukherjee, Hisao Nakanishi and Norman H. Fuchs}
\address{Department of Physics, Purdue University, W. Lafayette, IN 47907}
\date{\today}
\maketitle
\begin{abstract}
We study the dynamical exponents $d_{w}$ and $d_{s}$ for a particle
diffusing in a disordered medium (modeled by a percolation cluster), from
the regime of extreme disorder (i.e., when the percolation cluster
is a fractal at $p=p_{c}$) to the Lorentz gas regime when the cluster
has weak disorder at $p>p_{c}$ and the leading behavior is standard
diffusion. A new technique of relating the velocity autocorrelation
function and the return to the starting point probability to the
asymptotic spectral properties of the hopping transition probability
matrix of the diffusing particle is used, and the latter is numerically
analyzed using the Arnoldi-Saad algorithm.  We also present evidence for
a new scaling relation for the second largest eigenvalue in terms of the size
of the cluster, $|\ln{\lambda}_{max}|\sim S^{-d_w/d_f}$, which provides
a very efficient and accurate method of extracting the spectral dimension
$d_s$ where $d_s=2d_f/d_w$.
\end{abstract}

\section{Introduction}
\label{sec:intro}
Random walks have held the interest of physicists, engineers and
mathematicians for a long time now, partly due to their ability
to model a wide variety of problems which can be described at some
level by transition in states which are governed by probabilistic laws.
For example, random walks have been used extensively to model the
motion of a particle in diffusive processes through solids (ordered or
disordered), liquid or gas, including transport problems like
electrical conduction \cite{random,straley}, and in describing certain
conformational properties of macromolecules \cite{degennes,des}.

In this paper we present an approach to studying random walks which is
an alternative to the more common one where the individual random walk
is pursued either by simulation or in some other way.  This is the method
of spectral analysis of the corresponding transition
probability matrix. This method allows us to relate the various dynamical
quantities of the walk, such as the velocity autocorrelation function,
$\langle{\bf v}(t)\cdot{\bf v}(0)\rangle$, the return to the starting point
probability after $t$ steps, $P(t) $, the mean square displacement, and
the acceleration autocorrelation function, to the asymptotic spectral
properties of the transition probability matrix ${\bf W}$ (defined below).
With the development of the various numerical methods of diagonalizing
large matrices, this approach can serve as a very powerful tool to
solve for these dynamical quantities.

In this work, we model a random medium by a percolation cluster, where
sites on a lattice are occupied with probability $p$ (otherwise empty)
and the occupied nearest neighbor sites are considered to be connected
\cite{stauffer}. It is well known that there is a critical threshold
$p_c$ where the {\em incipient} infinite cluster is a disordered fractal
\cite{mandelbrot}. Such a cluster will serve as an example of a random
medium with a critically correlated disorder.  We will also confine
our attention for the moment to the random walk which takes
place on the (critical or noncritical) percolation cluster according to
the so-called blind or myopic ant rule. The blind ant moves to an occupied
nearest neighbor in one time step with equal probability of $ 1/z $
(where $ z $ is the coordination number of the underlying lattice)
and it remains at the same site with the probability equal to $(z-z_{i})/z$
(where $ z_{i} $ is the number of occupied
nearest neighbors of that particular site). On the other hand,
the myopic ant never stays at the same site but always moves to an occupied
nearest neighbor with equal probability $1/z_{i}$ at each time step
\cite{havlin}.

The transition probability matrix ${\bf W}$ is defined as the matrix with
its {\em ij}-th element equal to the probability that a particle at site
$j$ hops to site $i$ in the next time step.  Both the blind and myopic
ants can be described by a matrix ${\bf W}$ which is a Markov matrix
\cite{vankampen} as each element is equal to or greater than zero
and the sum of all the elements of each column is equal to one.
Such a matrix describes a {\em Markov chain}, which is a one-dimensional,
discrete Markov process.

In this paper we will mainly concentrate on the two important quantities
for the random walker, $\langle {\bf v}(t)\cdot{\bf v}(0) \rangle$ and
$P(t)$. At $p=p_{c}$ blind and myopic ants are known to exhibit a
nondiffusive leading behavior called {\em anomalous} diffusion
\cite{gefen}. In this case, the walk is often characterized by a
{\em walk dimension} $d_w$ where its mean square displacement after $t$
steps $\langle R(t)^2 \rangle$ is given by
\begin{equation}
\langle {\bf R}(t)^{2} \rangle \sim t^{2/d_{w}} .
\end{equation}
The deviation of $d_w$ from $2$ is a measure of the nonstandard nature
of the random walk.

In a stationary ensemble (that is, where the correlation functions depend
only on the relative time difference and not on the absolute times),
the second derivative of $\langle {\bf R}(t)^{2} \rangle $ is proportional
to $ \langle {\bf v}(t)\cdot{\bf v}(0)\rangle $. So under these conditions
$ \langle {\bf v}(t)\cdot{\bf v}(0) \rangle $ is normally expected to obey
the following power law in the long time limit:
\begin{equation}
\langle {\bf v}(t)\cdot{\bf v}(0) \rangle \sim t^{2/d_{w}-2}.
\end{equation}
Such a behavior is common for the blind ant and also for the myopic ant
on so-called nonbipartite lattices at least asymptotically (although
the leading behavior in the velocity autocorrelation for the myopic ant
on bipartite lattices do not derive from that of $\langle R(t)^2 \rangle$,
thus providing exceptions to this behavior).
A lattice (or a cluster) is called bipartite if it can be partitioned into
two equivalent classes of sites where all connectivity is between sites
of different classes.

The long-time behavior of the probability $P(t)$ that a random walker
returns to the starting point at time $t$ is also generally a power law,
where
\begin{equation}
P(t) \sim t^{-d_{s}/2},
\end{equation}
and $ d _{s} $ is the spectral dimension.  For a standard diffusive
behavior, $d_s$ is equal to the spatial dimensionality $d$, while
on a fractal, it is generally a fractional number different from $d$.
Thus one can see that the two exponents $d_w$ and $d_s$ characterize
the important quantities $\langle {\bf v}(t)\cdot{\bf v}(0) \rangle$
and $P(t)$ for the diffusion process in a fractal medium.

As $p$ is increased above $ p_{c} $, the percolation cluster becomes
nonfractal, and the particle on it starts to diffuse normally at large
length scales.  In this case, we have a random walk with fixed scatterers
placed with weakly correlated disorder, and thus we may describe this
regime as the Lorentz gas regime \cite{alder,ernst}.
In a Lorentz gas, $ \langle {\bf v}(t) \cdot {\bf v}(0) \rangle $
generally exhibits a {\em long-time tail}, obeying the following power law:
\begin{equation}
\langle {\bf v}(t) \cdot {\bf v}(0) \rangle \sim t^{-d/2-1} ,
\end{equation}
and $P(t)$ scales as in normal diffusion,
\begin{equation}
P(t) \sim t^{-d/2} ,
\end{equation}
where $d$ is the Euclidean dimension of the cluster.

Previously  diffusion in the two regimes (the anomalous diffusion regime
and the Lorentz gas regime) were often studied by using very
different techniques.  We show in this paper, the method of spectral
analysis of the hopping transition probability matrix provides an elegant
way of studying both regimes using the same method.

This work is organized as follows.
In Section\ \ref{sec:pro}, we review and summarize important properties
of the matrix ${\bf W}$, and in Section\ \ref{sec:spectrum} we discuss the
significance of the eigenspectrum of ${\bf W}$.  The particular significance
of the subdominant eigenvalue $\lambda_{max}$ is shown in
Section\ \ref{sec:scaling}.  In Section\ \ref{sec:saad}, the Arnoldi-Saad
approximate diagonalization scheme is discussed, and the main numerical
results of this work are given in Section\ \ref{sec:results}.
A summary of our findings is presented in Section\ \ref{sec:summary}.
A preliminary account of this work was given \cite{nakanishi} where
some of the present results were already reported.  For completeness,
we also reproduce those results here with more details, together with new
ones particularly on the scaling of the subdominant eigenvalue.

\section{Hopping transition probability matrix}
\label{sec:pro}
Since the transition probability matrix ${\bf W}$ contains all the information
about the geometry of the underlying substrate on which the random walk takes
place and also the kinematics of the walk itself, all the characteristics of
the walk can be extracted from this matrix. Because of the importance of
${\bf W}$ we enumerate in this section some properties of ${\bf W}$,
especially of
its eigenvalue spectrum.  Much of the information is hidden in its eigenvalue
spectrum as will be elucidated later. Some of the properties discussed here
are already well known but are reproduced here for completeness, while
others are not commonly appreciated despite their potential usefulness.

\subsection{Basic properties}
By definition, Markov matrices must have the following two properties.
First, for all $i$ and $j$,
\begin{equation}
W_{ij} \geq 0
\end{equation}
and second,
\begin{equation}
\sum_{i} W_{ij} = 1 .
\end{equation}
The second property ensures the presence of the eigenvalue 1 with at least
one normalized left eigenvector having all components positive and
of the same magnitude.  There is also a corresponding normalized right
eigenvector with eigenvalue 1.
This right eigenvector in fact represents the equilibrium
probability distribution of the diffusing particle.
Thus, for a connected cluster (with no isolated, unreacheable part),
the corresponding eigenvector of ${\bf W}$ can be made positive-definite.

It is well known that the  eigenvalues of the Markov matrix satisfy
\begin{equation}
\label{eq:prop1}
|\lambda|\leq 1 .
\end{equation}
For completeness sake a proof is given here.  Let $ u_{j} $ be the left
eigenvector of ${\bf W}$ corresponding to the eigenvalue $\lambda$, then
\begin{equation}
\sum_{j}W_{ij}u_{j}=\lambda u_{i} .
\end{equation}
Taking the modulus of the above equation, we get,
\begin{equation}
| \sum_{j} W_{ij} u_{j} | = | \lambda | | u_{i} |,
\end{equation}
\begin{equation}
\sum_{j} W_{ij}| u_{j} | \geq |\lambda| | u_{i} | .
\end{equation}
Now summing over the index $i$ in the above equation, we get
\begin{equation}
\sum_{j}| u_{j}| \sum_{i}W_{ij}
\geq |\lambda| \sum_{i}| u_{i}| ,
\end{equation}
\begin{equation}
\sum_{j}| u_{j} | \geq | \lambda|\sum_{i} | u_{i}| .
\end{equation}
We have made use of the second property of the Markov matrix.  The result
Eq.\ (\ref{eq:prop1}) then follows immediately.

\subsection{Detailed and extended detailed balance}
The transition matrix ${\bf W}$ is said to satisfy the detailed balance
condition if its elements satisfy the following condition for all $(i,j)$:
\begin{equation}
W_{ij}\rho^{e}_{j}=W_{ji}\rho^{e}_{i} ,
\end{equation}
where $ \rho^{e}_{j} $ is the $j$-th element of the equilibrium probability
distribution, i.e., the normalized, positive-definite right eigenvector of
${\bf W}$ for the eigenvalue 1. We now show that ${\bf W}$
satisfying the detailed balance condition can be transformed
into a symmetric matrix by a similarity transformation,
thus proving that all eigenvalues of ${\bf W}$ are real.

A proof goes as follows:
Taking the square root of the above equation we get
\begin{equation}
(W_{ij})^{1/2}(\rho^{e}_{j})^{1/2}=(W_{ji})^{1/2}(\rho^{e}_{i})^{1/2} .
\end{equation}
Now, let the matrix ${\bf A}$ be diagonal with the $i$-th diagonal element
equal to $({\rho_i^e})^{1/2}$. Then its inverse ${\bf A}^{-1}$ is also
diagonal with its $j$-th diagonal element equal to $({\rho_j^e})^{-1/2}$.
We will show that the matrix ${\bf A}^{-1}{\bf W}{\bf A}$ is
symmetric.  Let us consider the $ij$-th element,
$(\rho_{i}^{e})^{-1/2} W_{ij} (\rho_{j}^{e})^{1/2}$, of this matrix:
\begin{eqnarray}
 \nonumber
(\rho^{e}_{i})^{-1/2}W_{ij}(\rho^{e}_{j})^{1/2}
 & = & (\rho^{e}_{i})^{-1/2}(W_{ij})^{1/2}(W_{ij})^{1/2}(\rho^{e}_{j})^{1/2} \\
 \nonumber
 & = & (\rho_{j}^{e})^{-1/2}(W_{ji})^{1/2}(W_{ji}\rho_{i}^{e})^{1/2} \\
 & = & (\rho_{j}^{e})^{-1/2}W_{ji}(\rho_i^{e})^{1/2} ,
\end{eqnarray}
which is just the $ji$-th element, thus completing the symmetry proof.

The matrix ${\bf W}$ is said to satisfy the extented detailed balance
condition of order $n$ if it satisfies the following property for all $(i,j)$:
\begin{equation}
(W^{p})_{ij}\rho^{e}_{j}=(W^{p})_{ji}\rho^{e}_{i}
\end{equation}
for $p=n$ but not for $p=n-1$. From the previous argument on the detailed
balance condition we can say that all eigenvalues of ${\bf W}^{n}$ are real
if ${\bf W}$ satifies this condition. But ${\bf W}$ will in general have
complex eigenvalues in conjugate pairs (as ${\bf W}$ itself is real).
Thus all complex eigenvalues of ${\bf W}$ are such that $\lambda^{n}$ are
real, as the eigenvalues of ${\bf W}^{n}$ are just the $n$-th power of those
of ${\bf W}$.  Therefore, all the complex eigenvalues of ${\bf W}$ which have
the same magnitude converge to either a single real eigenvalue for
${\bf W}^{n}$ (which is $ \lambda^{n} $) or to
the two values $\pm|\lambda^{n}|$. Since the
eigenvectors of ${\bf W}$ and ${\bf W}^{n}$ are the same, all the linearly
independent eigenvectors corresponding to the different eigenvalues
having the same magnitude will have the same one or two eigenvalues,
producing a degeneracy of order between $n/2$ and $n$ for ${\bf W}^n$.

The detailed balance condition is satisfied by most models of diffusion
(such as the blind and myopic ant random walks considered here), but
it is possible to violate it by considering, e.g., {\em directed} walks.
A particularly simple example is a random walk taking place on a unit square
but with a preference for moving clockwise.  In this case, the
detailed balance is not obeyed, but the extended detailed balance {\em is}
obeyed at the 4th order.  More involved examples may be found when the
kinetic rules of diffusion induce the formation of {\em vortices},
reminiscent of what actually happens in turbulence.

\subsection{Oscillations in autocorrelations}
The position autocorrelation function can be expanded in terms of the real
(positive and negative) and complex eigenvalues in the following way:
\begin{eqnarray}
 \nonumber
\langle{\bf r}(n)\cdot{\bf r}(0)\rangle =
\sum_{\lambda_{+}}a(\lambda_{+})\lambda_{+}^{n}
 & + & (-1)^{n} \sum_{\lambda_{-}}a(\lambda_{-})\lambda_{-}^{n} \\
 & + & \sum_{\lambda_{c},\omega} 2 a(\lambda_{c},\omega)
\cos[\omega n + \phi(\lambda_{c},\omega)]\lambda_{c}^{n} ,
\end{eqnarray}
where $\lambda_+$ and $-\lambda_-$ are real, positive and negative
eigenvalues respectively and $\lambda_c e^{i\omega}$ denotes a complex
eigenvalue with modulus $\lambda_c$.
The last sum is over the complex eigenvalues and thus relevant only for
the matrices ${\bf W}$ which do not satisfy the detailed balance condition.
The cosine term arises in this case because the complex roots come in
conjugate pairs.  This term, if present, gives rise to oscillations in
the position autocorrelation function, different from the even-odd oscillation
that stems from the negative real eigenvalues.  This oscillation is persistent
(shows up even in the long time limit) if the
prefactor of the cosine term $\lambda_{c}^n$ is equal to one.
For $\lambda <1$, the oscillation is exponentially damped.

Thus ${\bf W}$ (if it does not satisfy the detailed balance condition) can
have both persistent and transient oscillations in the position autocorrelation
function of period larger than two. Now, if ${\bf W}$ satisfies the extended
detailed balance of order $n$, then the autocorrelations will have oscillations
of period at most $2n$, again, either persistent or transient depending
on the modulus of the eigenvalue causing the oscillation.  In this way,
the order of the extended detailed balance and the period of oscillation
in the autocorrelation are closely related.  This is intuitively
acceptable when we consider the case of vortices for a typical
example of ${\bf W}$ satisfying the extended detailed balance condition.

\subsection{Myopic and blind ant}
In light of the above discussion, let us now concentrate on the properties
of ${\bf W}$ for the myopic and blind ants specifically.
Blind ant in a disordered medium has a symmetric ${\bf W}$ as all the
nonzero off-diagonal elements are equal to the same value, $1/z$
(where $z$ is the coordination number of the lattice). Thus all its
eigenvalues are real. It does not have an eigenvalue equal to $-1$, so its
position autocorrelation function can only have transient, even-odd
oscillation.

On the other hand, ${\bf W}$ for the myopic ant is not symmetric as
the hopping probability depends on the location of the random walker,
that is, it is equal to $1/z_{i}$, where $z_{i}$ is the number of
occupied nearest neighbors of the particular site.  However, it does
satisfy the detailed balance condition, and thus can be transformed into
a symmetric matrix by a similarity transformation as seen before.
Thus the myopic ant also can have at most an even-odd oscillation.
However, such an oscillation can be persistent since an eigenvalue $-1$
is possible.  This occurs when the cluster is bipartite.
This is because the myopic ant on a bipartite cluster have
eigenvalues in pairs of $\lambda $ and $-\lambda$,
as can be seen easily by converting an eigenvector corresponding to
eigenvalue $\lambda $ into one corresponding to eigenvalue $-\lambda$
by simply making the components corresponding to the sites of one of the
subclusters opposite in sign.  Myopic ants on a nonbipartite cluster,
however, can only have a transient, even-odd oscillation.

\section{Eigenspectrum of ${\bf W}$ and dynamics of the walk}
\label{sec:spectrum}
Now we embark on the task of understanding the eigenspectrum of ${\bf W}$
and building up the connection between the spectrum and
the characteristic functions of the walk in the time domain (such as
the velocity autocorrelation function and the return to the starting-point
probability).

\subsection{Density of eigenvalues $ n(\lambda) $}
In this work, let us define the density of eigenvalues $n(\lambda )$ as
the number of eigenvalues of ${\bf W}$ normalized by the range of $\lambda$
as well as by the number of sites $S$ in the cluster.  (This differs from
the convention used in \cite{nakanishi} by a factor of $S$.)
The densities $n(\lambda)$ for the blind and myopic ants
show some interesting common characteristics. First, at the critical
dilution $p=p_c$, the eigenvalues accumulate as $\lambda \rightarrow 1$,
which results in a power law increase of $n(\lambda )$ as
$\ln \lambda \rightarrow 0$.  As $p$ is increased from its critical value,
the piling up of the eigenvalues diminishes and the value of $n(\lambda )$
averaged over small regions of $\lambda$
either flattens (in $d=2$) or actually decreases (in $d=3$) as
$\ln \lambda \rightarrow 0$.
This is illustrated in Fig.\ \ref{fig1} where $n(\lambda )$ for the
myopic ant on a small square lattice is plotted
for $p=0.593 \approx p_{c}$ and $p=0.8 > p_c$.  Second, the eigenvalue
spectrum shows sharp peaks at certain characteristic values of $\lambda$
in a region far from the asymptotic limit of $\lambda \rightarrow 1$.
This is better illustrated in Fig.\ \ref{fig2} for the blind ant on
a small lattice.  These observations are disucssed in more details below.

First, we recall the connection between the density $n(\lambda )$ and
the probability $P(t)$ that the diffusing particle returns to the
starting point at time $t$.  The mean probability $P(t)$ can be expressed
in terms of the corresponding probability $P_i(t)$ for a given starting
point $i$ as
\begin{equation}
P(t)=\sum_{i} P_i(t) \rho_o(i) ,
\end{equation}
where $\rho_o(i)$ is the probability for the particle to be at site $i$
initially.  When the initial distribution is the equilibrium one (as we
normally assume), this probability is uniformly equal to $1/S$ (where $S$
is the cluster size) for the blind ant while, for the myopic ant, it
fluctuates from site to site according to the local coordination number.
In either case, $P_i(t)$ is given by
\begin{equation}
P_i(t) = {\bf e}_{i}^{T} {\bf W}^t {\bf e}_{i} ,
\end{equation}
where ${\bf e}_{i}$ is the column vector whose components are all zeros except
the $i$-th component which is one and
$ {\bf e}_{i}^{T} $ is the corresponding row vector of $ {\bf e}_{i} $.

Below we assume the blind ant for simplicity, but the final result is
essentially the same for the myopic ant as will be noted.
Thus,
\begin{equation}
\label{eq:trace}
P(t) = \mbox{Tr} {\bf W}^t / S = \sum_{i}{\lambda_{i}}^t /S .
\end{equation}
Since $-1$ is not an eigenvalue for the blind ant and there is a buildup
of eigenvalues only near $ \lambda = 1 $ (for $p=p_{c}$),
and since we are only interested in the long time limit where the
dominant contribution comes from the eigenvalues near one, we can neglect
the contribution of the negative eigenvalues.
Thus summing the eigenvalues in the above equation only over $ \lambda > 0$
and by taking the inverse Laplace transform, we get
\begin{eqnarray}
 \nonumber
\frac{1}{2\pi i} \int_{\Gamma} e^{Ez} \mbox{Tr} {\bf W}^z / S & \approx &
 \sum_{\lambda >0} \frac{1}{2\pi}\int_{-\infty}^{\infty}dt
               e^{i(E-|\ln\lambda|)t} / S \\
 & = & \sum_{\lambda > 0}\delta(E-|\ln\lambda|) /S ,
\end{eqnarray}
where Eq.\ (\ref{eq:trace}) has been continued analytically to the complex
plane and $\Gamma$ is along the imaginary axis.

The quantity on the right is the density of eigenvalues as normalized in
$E$ (i.e., in the space of $\ln \lambda$), but for small $E$ (or for
eigenvalues near one), it reduces to $n(\lambda)$.  Thus we see that $P(t)$
is the Laplace transform of the (normalized) $n(\lambda)$, at least for
large $t$ where the neglect of $\lambda < 0$ is unimportant.
Although the above discussion has been for the blind ant, the calculation
for the myopic ant is also very similar.  In the latter case, $P(t)$ will
be bounded by the Laplace transform of $n(\lambda )$ multiplied by two
different coefficients.

Since $P(t)$ for the ants in general decays for large $t$ following
a power law\cite{havlin},
\begin{equation}
P(t) \sim t^{-d_{s}/2} ,
\end{equation}
the Laplace relationship implies for both blind and myopic ants,
\begin{equation}
n(\lambda) \sim {| \ln\lambda|}^{d_{s}/2-1} .
\end{equation}
Similarly for $p>p_{c}$, $P(t)$ decays as
\begin{equation}
P(t) \sim t^{-d/2} ,
\end{equation}
where $ d $ is the Euclidean dimension of the lattice, leading to
\begin{equation}
n(\lambda) \sim |\ln\lambda|^{d/2-1} .
\end{equation}

\subsection{Autocorrelation function}
The position autocorrelation function of a diffusing particle can be
expressed in the following way:
\begin{equation}
\langle {\bf r}(t)\cdot{\bf r}(0)\rangle =
  {\bf X^T} {\bf W}^t {\bf X_o} + {\bf Y^T} {\bf W}^t {\bf Y_o}
                              + {\bf Z^T} {\bf W}^t {\bf Z_o} ,
\label{eq:pv}
\end{equation}
where ${\bf X^T}$, ${\bf Y^T}$, and ${\bf Z^T}$ are the row vectors
whose components are the $x$, $y$, and $z$ components of the corresponding
sites, respectively, while ${\bf X_o}$, ${\bf Y_o}$, and ${\bf Z_o}$ are
the column vectors whose $i$-th components are equal to $\rho_i X_i$,
$\rho_i Y_i$, and $\rho_i Z_i$, respectively,
where ${\bf \rho}$ is the initial probability distribution for the random
walker.

The vectors ${\bf X }$, ${\bf Y }$, ${\bf Z }$ can be expanded in terms of
the eigenvectors of the transition matrix ${\bf W}$.  For example, we have
\begin{equation}
{ \bf X } = \sum_{\lambda} b^x_{\lambda} { \bf u }_{\lambda} ,
\label{eq:coeff1}
\end{equation}
where $ b^x_{\lambda} $ is the expansion coefficient and ${\bf u }_\lambda$
is the normalized right eigenvector with the eigenvalue $\lambda$.
Similarly the vectors ${\bf X_o} $, ${\bf Y_o}$, ${\bf Z_o}$ can be
expressed in terms of the normalized right eigenvectors of ${\bf W}$.
For example, we can write
\begin{equation}
{\bf X_o} = \sum_{\lambda}c^x_{\lambda}{\bf u }_{\lambda},
\label{eq:coeff2}
\end{equation}
where $ c^x_{\lambda} $ is the expansion coefficient.

Thus Eq.\ (\ref{eq:pv}) can be expressed as
\begin{equation}
\langle {\bf r}(t)\cdot{\bf r }(0)\rangle =
   \sum_{\lambda} (b^x_{\lambda}c^x_{\lambda}
   + b^y_{\lambda}c^y_{\lambda}
   + b^z_{\lambda}c^z_{\lambda}) {\lambda}^t ,
\end{equation}
noting that the normalized eigenvectors ${\bf u}_{\lambda}$ are orthogonal
to each other.  Further denoting
\begin{equation}
a_{\lambda} \equiv \vec{b}_{\lambda} \cdot \vec{c}_{\lambda} ,
\label{eq:coeff3}
\end{equation}
where the vector sign and dot product refer to the coordinate space,
we can write
\begin{equation}
\langle{\bf r}(t)\cdot{\bf r}(0)\rangle
 = \sum_{\lambda} a_{\lambda} {\lambda}^t .
\end{equation}

Though we have calculated the position autocorrelation function here,
other correlation and autocorrelation functions can also be calculated
in a similar way.  This includes the mean square displacement
$\langle ({\bf r}(t)-{\bf r}(0))^2 \rangle$ and position-velocity cross
correlation $\langle{\bf r}(t)\cdot{\bf v}(t)\rangle$, as well as the
velocity autocorrelation function.  The latter has been shown \cite{fuchs}
to be related to the eigenspectrum of ${\bf W}$ in the following way:
\begin{equation}
\langle {\bf v}(t)\cdot{\bf v}(0) \rangle
 = - \sum_{\lambda} a_{\lambda} (\lambda -1)^2 {\lambda}^t .
\label{eq:vv}
\end{equation}

Since the eigenspectrum of the myopic ant on a bipartite cluster contains
both the eigenvalues 1 and $-1$, after {\em all} the transients have died out,
its position autocorrelation function behaves as
\begin{equation}
\langle {\bf r}(t)\cdot{\bf r}(0) \rangle \sim a_{1} + a_{-1}(-1)^n .
\end{equation}
That is, the position autocorrelation has an undamped oscillation about the
value $a_1$. If the cluster is nonbipartite or if the ant is a blind ant,
then there is no eigenvalue $-1$, and thus there will be no undamped
oscillations.  It can easily be seen from Eq.~(\ref{eq:vv}) that the
corresponding velocity autocorrelation for the myopic ant on a bipartite
cluster also has an undamped oscillation.  However, because of the factor
$(\lambda -1)^2$, this persistent oscillation is symmetric about
zero as shown in \cite{jacobs}. Again, if the cluster is not bipartite or
if the ant is a blind ant, then such an undamped oscillation will not exist.

Since the long time behavior of autocorrelations for the blind ant and
the myopic ant on a nonbipartite cluster are (normally) dominated by
the eigenspectrum near one, the summation over the eigenvalues
in Eq.\ (\ref{eq:vv}) can be restricted to only the positive ones.
For the myopic ant on a bipartite cluster, however, there is an accumulation
of eigenvalues symmetrically about $\lambda = \pm 1$.  Thus in this case,
the factor $(\lambda -1)^2$ actually makes the eigenvalues near $-1$
dominate over those near one, and the summation can be replaced by the one
over only the negative eigenvalues.

Therefore, in the long time limit, the velocity autocorrelation can be
written as
\begin{equation}
\langle {\bf v}(t)\cdot{\bf v}(0) \rangle
 = - \sum_{\lambda \in \Lambda} a_{\lambda} (\lambda -1)^2 {\lambda}^t ,
\end{equation}
where $\Lambda$ is the set of positive eigenvalues near one for the blind
ant and the myopic ant on a nonbipartite cluster (Case A)
and the set of negative eigenvalues near $-1$ for the myopic ant
on a bipartite cluster (Case B).
Converting the summation in the above equation into an integral,
\begin{equation}
\langle{\bf v}(t)\cdot{\bf v }(0) \rangle
 \approx - \int_{\Lambda} d\lambda \pi (\lambda ) {\lambda}^{t} ,
\end{equation}
where $ \pi(\lambda) $ is
\begin{equation}
\pi(\lambda) = n(\lambda) a_{\lambda} S (\lambda-1)^2 ,
\end{equation}
where $ n(\lambda) $ is the density of eigenvalues per cluster site.

This equation can be further cast in the form of the Laplace transform
in the long time limit,
\begin{equation}
|\langle{\bf v}(t)\cdot{\bf v }(0) \rangle |
 \approx \int_{0}^{\infty} du {\rm e}^{-ut} \pi (\lambda ) ,
\end{equation}
where $u=|\ln \lambda |$ for Case A (blind ant in general and myopic ant
on nonbipartite cluster), and $u=|\ln |\lambda ||$ for Case B
(myopic ant on a bipartite cluster).  This establishes the Laplace
relationship between $\pi (\lambda )$ and the velocity autocorrelation.
Therefore the power law behavior of the velocity autocorrelation
\begin{equation}
|\langle {\bf v}(t) \cdot {\bf v}(0) \rangle | \sim {t}^{-y}
\end{equation}
implies a corresponding power law behavior in $\pi (\lambda )$,
\begin{equation}
\pi(\lambda) \sim u^{y-1} ,
\end{equation}
with $u$ identified appropriately as above.
Thus, by measuring the exponent $y-1$ for the function $\pi (\lambda )$,
we can deduce the corresponding asymptotic behavior of the velocity
autocorrelation function.

The exponent $y$ appearing in the velocity autocorrelation depends on
the two cases \cite{nakanishi,fuchs},
\begin{equation}
y =\left\{\begin{array}{ll}
    2 - 2/d_w , & \mbox{(Case A, at $p_c$)} \\
    d_s /2 ,    & \mbox{(Case B, at $p_c$)}
          \end{array}
   \right.
\end{equation}
for $p=p_c$, and
\begin{equation}
y =\left\{\begin{array}{ll}
    d/2 + 1 ,   & \mbox{(Case A, $p>p_c$)} \\
    d/2 ,       & \mbox{(Case B, $p>p_c$)}
          \end{array}
   \right.
\end{equation}
for $p>p_c$. In Case A, this exponent characterizes the decay of the
autocorrelation which is always negative for long times (cage effect),
while in Case B, it describes the decrease in the envelope amplitude
of the even-odd oscillation of the autocorrelation function.

In the latter case, however, the mean of the autocorrelation function
from the two consecutive even-odd steps (rather than the envelope) does
have the same decay behavior as in the first case.  Thus, the behavior
of the center line for the oscillating velocity autocorrelation for the
myopic ant on bipartite clusters can be obtained from the eigenspectrum
near 1, allowing it to be included with Case A. In fact, some of the
numerical results we present later demonstrate that such behavior leads
to exactly the same exponent $y$ as in Case A.

\subsection{Local structure in the eigenvalue spectrum}
The eigenspectrum shows sharp peaks at certain characteristic values of
$ \lambda $. The $ \lambda $ at which these peaks occur depends upon the type
of lattice and the kind of ant, indicating that the peaks are structures
corresponding to small length scales.  For example, the blind ant on
the square lattice shows strong characteristic peaks
at $\lambda =0.5$ and 0.75.

The presence of high degeneracies for these eigenvalues
can be explained by looking at their corresponding eigenvectors.
For example, the eigenvectors for the eigenvalue 0.5 reveal
that the components corresponding to most of the sites are zero and
those with nonzero values are arranged in pairs of equal
magnitudes but opposite signs. In fact, they involve the diagonal
arrangement of two sites of a unit square having amplitudes of
opposite signs with the remaining two sites of the square having
zero amplitudes.  In some cases, there are additional pairs of sites
with amplitudes of opposite signs hanging off of such arrangements.
Similarly, when one inspects the components of the eigenvectors for the
eigenvalue 0.75, they are seen to involve one, two or three dangling
ends which are connected to the main part of the cluster through
the same site which has a zero amplitude.  Again, in some cases, there
are additional pairs of sites with nonzero amplitudes attached to
such arrangements. Some examples of these arrangements are illustrated
in Fig.\ \ref{fig3}.

These local structures connecting a small number of sites are simple
enough to be present in large numbers for essentially all clusters
of appreciable size.  Thus they give the sharp peaks seen in the
eigenspectra as in Fig.\ \ref{fig1}, but do not contribute much
to the long time behaviors of diffusion on the structure.

\section{Scaling for the subdominant eigenvalue}
\label{sec:scaling}
As we have discussed, at $p=p_c$ the eigenvalues of ${\bf W}$ accumulate
near one though the eigenvalue 1 itself (if it exists) remains isolated
from this build up. This behavior near the eigenvalue 1 is mirrored in
the region near $-1$ for the myopic ant on bipartite cluster.
Thus for any finite cluster there is a sharp drop from the maximum eigenvalue
1 to the subdominant eigenvalue $\lambda_{max}$. This gap decreases to
zero as the cluster size increases to infinity.

A simple scaling argument for the cluster size dependence of $\lambda_{max}$
can be given as follows.  In the Laplace relationship between the time
domain quantities and the corresponding functions of $\lambda$, the variables
$t$ and $|\ln \lambda|$ are the conjugate variables.  Thus,
$|\ln \lambda|^{-1}$ corresponds to a time scale.  In particular,
$|\ln \lambda_{max}|^{-1}$ corresponds to the time scale associated with
the eigenmode with the eigenvalue $\lambda_{max}$.  In terms of the
mapping between the random walk and {\em scalar} elasticity problems
\cite{alexander}, the vibrational energy $E$ of a mode
corresponds to $|\ln \lambda |$; thus again $|\ln \lambda |^{-1}$
corresponds to a time scale.

Since the subdominant mode for $\langle R(t)^2 \rangle$ (which is actually the
{\em dominant} one for the velocity and acceleration autocorrelations)
should reflect the geometry and size of the
whole cluster, the time scale $\ln \lambda_{max}$ should correspond to the
time it takes for a diffusing particle to {\em just} sample the entire cluster
of size $S$.  The latter time scales as $S^{d_w/d_f}$.  Therefore, we expect
\begin{equation}
|\ln\lambda_{max}| \sim S^{-d_w/d_f} ,
\end{equation}
where the exponent $d_w/d_f$ can be replaced by $2/d_s$ for the critical
percolation cluster \cite{alexander}. This way of calculating $d_s$ is
very efficient since
it only requires the accurate determination of one eigenvalue (the second
largest one) in the entire eigenspectrum. The numerical confirmation of
this power law is described in the section on the numerical results.

We now propose a more general scaling relation which includes the region
near but greater than $p_{c}$ as follows:
\begin{equation}
|\ln\lambda_{max}| = S^{-d_w/d_f} f((p-p_{c}) S^\sigma) ,
\label{eq:general}
\end{equation}
where $f(x)$ is a universal scaling function and the exponent
$\sigma = 1/(d_f \nu )$ is the one appearing in the usual percolation
cluster size scaling \cite{stauffer}.

In order to satisfy the known asympototic behavior for $p>p_c$, we
assume that $f(x)$ satisfies the following limiting behavior:
\begin{equation}
f(x) =\left\{\begin{array}{ll}
               \mbox{constant}, & x\rightarrow 0 \\
               x^{z},          & x\rightarrow \infty
             \end{array}
      \right.
\label{eq:pho}
\end{equation}
We see that in the $ x\rightarrow 0 $ limit we regain the previous relation,
and the exponent $z$ can be determined by demanding that the
$ x\rightarrow \infty $ limit leads to the normal diffusive behavior
where $d_w =2$ and $d_f =d$, the Euclidean dimension.  Thus, in this limit
we expect
\begin{equation}
|\ln \lambda_{max}| \sim {S}^{-2/d}
\label{eq:vibra}
\end{equation}
For this to be true, we must then have
\begin{equation}
z=(d_{w}/d_{f} - 2/d)/\sigma .
\end{equation}

Since the cluster is {\em ordered} for $p>p_c$, the scalar elasticity
analog would be the phonon problem. In this case the energy of the
phonons $E \sim k^{2} \sim l^{-2}$, where $l$ is a length-scale of
the phonon, and since the cluster has the dimensionality $d$,
$l_{max} \sim S^{1/d}$.  Thus, we again see that
$E_{min} \sim |\ln \lambda_{max}| \sim {S}^{-2/d}$ as above.
Numerical results supporting the above finiste size scaling relation for
$|\ln\lambda_{max}|$ are also given in the section on numerical results.

\section{Arnoldi-Saad approximate diagonalization method}
\label{sec:saad}
In this work we have used the Arnoldi-Saad algorithm \cite{saad} in order
to approximately diagonalize the transition probability matrix ${\bf W}$.
The main advantage of using this algorithm in diagonalizing a matrix is
that it allows the very accurate determination of a subset of the eigenvalues
near the maximum in the spectrum.  This is convenient since we are interested
in the asymptotic long-time behavior of the diffusion process which are
determined by the eigenvalues near the maximum.  In particular, we are
mainly interested in the asymptotic power law behavior of the density of
eigenvalues $n(\lambda )$ and the function $\pi (\lambda )$ as well as
that of the subdominant eigenvalue $\lambda_{max}$ as discussed above.
Both because the thermodynamic limit requires a large system and because
the eigenvalues approach 1 more closely for larger size of the cluster,
the accurate determination of the asymptotic behavior demands a large
cluster, and we need a large number of independent large clusters to obtain
true quenched disorder averages.  While the diagonalization of a large
number of large matrices is usually very time consuming and inefficient,
the Arnoldi-Saad algorithm affords a valuable reduction in the size of the
matrices to diagonalize without losing the required information.

It is interesting to note that the region of the spectrum most accurately
obtained by our approach is exactly the region of most interest. In previous
studies of the eigenspectrum of the fractal diffusion problem
\cite{nakayama}, the matrix diagonalized was not ${\bf W}$ but rather
a matrix constructed from the equation of motion for the analogous
scalar elasticity problem \cite{alexander}.  For the latter, the important
region of the eigenspectrum is near zero (recall the relation
$|\ln \lambda | \sim \omega^2$ where $\omega$ is the angular frequency
of vibration for the scalar elasticity problem). This is where an accurate
determination of the spectrum is most difficult by most methods.

A large amount of CPU time and memory is saved by using the Arnoldi-Saad
method which allows the reduction of the dimension of the original matrix
without compromising the needed accuracy of the results.
For instance a 5000$\times$5000 ${\bf W}$ matrix was reduced by this method
typically to one of size 300$\times$300, and the eigenvalues and eigenvectors
obtained were reliable to better than one part in $ 10^{5} $. This operation
required about 560 seconds of CPU time per cluster on a Kubota Pacific Titan
P3 mini-super computer or less than 30 seconds on a CPU of a Cray 2 super
computer, including the cluster generation itself.

The algorithm was described in the literature \cite{fuchs}, but we will
sketch it here for completeness.  In this algorithm, one starts with
an arbitrary normalized vector ${\bf u}_1$ of dimension $S$ where ${\bf W}$
has the dimension of $S \times S$. Then one chooses the subspace dimension
$m \leq S$ which is to be the dimension of the reduced upper Hessenberg
matrix ${\bf H}$ which will then be diagonalized essentially exactly.
The reduced matrix ${\bf H}$ is obtained recursively and at the same time
as a sequence of normalized basis vectors ${\bf u}_i$ ($i=2,...,m$)
in the following way:

First, the element $H_{11}$ is obtained as ${\bf u}_1^T {\bf W}{\bf u}_1$.
Second, the product ${\bf v}_2 \equiv H_{21} {\bf u}_2$ is calculated as
${\bf v}_2 = {\bf W}{\bf u}_1-H_{11}{\bf u}_1$.  We then choose
$H_{21} = \|{\bf v}_2\|$ to satisfy the normalization requirement
$\|{\bf u}_2 \|^2 =1$.  At the next iteration, we first obtain $H_{12}$ and
$H_{22}$ as ${\bf u}_1^T {\bf W}{\bf u}_2$ and ${\bf u}_2^T {\bf W}{\bf u}_2$,
respectively, and then, ${\bf u}_3$ and $H_{32}$ are obtained similarly
to the first iteration.  This process continues until all elements of
${\bf H}$ and all of the ${\bf u}_i$ are obtained.  In general, we have
\begin{eqnarray}
H_{ij} & = & {\bf u}_i^T {\bf W}{\bf u}_j  , \\
H_{j+1,j} {\bf u}_{j+1} & = & {\bf W}{\bf u}_j
- \sum_{i=1}^{j} H_{ij}{\bf u}_i  ,
\end{eqnarray}
where ${\bf u}_i$ ($i=1,2,...,m$) form an orthonormal set.

This procedure ensures \cite{saad} that the eigenvalues of
${\bf H}$ are approximate
eigenvalues of the original matrix ${\bf W}$ and that if ${\bf y}$ is
the eigenvector of ${\bf H}$ with the eigenvalue $\lambda$, then
the vector ${\bf z}$ defined by its components
$z_j \equiv \sum_{i=1}^m [u_i]_j y_i$ satisfies the orthogonality
\begin{equation}
({\bf W}-{\lambda}{\bf I}) {\bf z} \cdot {\bf u}_i = 0, \;\;\;\; i=1,2,...,m .
\end{equation}
This means that ${\bf z}$ is an approximate eigenvector of ${\bf W}$
to within the subspace spanned by the ${\bf u}_i$'s with the approximate
eigenvalue $\lambda$.

The approximation is better the larger $m$ is, and also the outer part of
the spectrum yields better accuracy \cite{saad}.  In practice, the
eigenvalues pile up at very high density near the outer edge of the spectrum
(for $p_c$), which requires ${\bf W^N}$ with a large $N$ (typically greater
than 100) to be used in place of ${\bf W}$ to separate the eigenvalues.
The accuracy of the eigenvalues and eigenvectors can be checked by a number
of criteria including one which uses the comparisons between the results
obtained by using different dimensions $m$ for the submatrix ${\bf H}$,
and also by using different $N$.

\section{Numerical Results}
\label{sec:results}
In this section, we summarize the numerical results for the density of
states per site $n(\lambda )$, the function $\pi (\lambda )$, and the
scaling of the subdominant eigenvalue $\ln \lambda_{max}$.

\subsection{Density of eigenvalues $ n(\lambda) $}
The density of states $n(\lambda)$ for the blind and myopic ants
has been calculated in two dimensions on the square lattice and
in three dimensions on the simple cubic lattice,
both with and without periodic boundary conditions.
In practice, $ n(\lambda) $ is computed by accumulating the eigenvalues in
logarithmic bins over a large number of independent cluster realizations
and then dividing by the number of clusters, the bin width, and the cluster
size.

In Fig.\ \ref{fig4}, $n(\lambda)$ at $p_c$ is plotted against
$|\ln\lambda|$ for the blind ant on nonperiodic clusters in $d=2$ and 3.
The averages over 5000, 250 and 900 clusters were taken for the cluster
sizes $S=400$, 800 and 5000, respectively in $d=2$,
while for $d=3$ the averages were taken over 250, 300 and 449
clusters for the cluster sizes $S=400$, 800 and 5000, respectively.
It is clear from the figure that the data for different cluster
sizes collapse into a single curve for both $d=2$ and 3, showing that finite
size effects for $n(\lambda)$ are not important. The approximate $p_c$ used
here is $p=0.593$ for the square lattice in $d=2$ and $0.312$ for the simple
cubic lattice in $d=3$.

In order to evaluate the exponent $d_{s}$ in $d=2$, a log-log plot
is produced for $n(\lambda)$ as a function of $|\ln\lambda|$ in
Fig.\ \ref{fig5} for the blind and myopic ants on the square lattice.
As seen clearly, the data for the blind ant on the periodic and nonperiodic
clusters collapse within the statistical errors on the same line of slope
$-0.35\pm0.01$, showing that boundary conditions do not affect the
exponent $d_{s}$.  The data for the periodic cluster have been collected
on 100$\times$100 grids and the average was taken over 400 clusters and
then normalized by the average size of the cluster.
For the nonperiodic cluster, the average was taken over 900 clusters
of 5000 sites. The data for the myopic ant on the periodic cluster,
on the other hand, yield a least squares fitted line of a slightly larger
slope $-0.37\pm0.01$. The dashed lines showing these fits had the
correlation coefficients greater than 0.999.
The data for the myopic ant were obtained on 100$\times$100 grids
and the average was taken over 250 clusters.
We believe that the small difference in the slopes between the blind and
myopic ants is a reflection of the inexact Laplace relationship between
$n(\lambda )$ and $P(t)$ for the case of the myopic ant.
The details of the exponent estimates extracted from these data along with
the values of the exponent quoted in a standard reference \cite{havlin}
are tabulated in Table\ \ref{table1}. The blind and myopic ants on triangular
lattice also had a similar slope but they are not shown in the figure to avoid
overcrowding.

The exponent $d_s$ for $d=3$ has similarly been extracted from log-log
plots of $n(\lambda)$ versus $|\ln\lambda|$ for the blind ant on the
periodic and nonperiodic simple cubic lattices and the myopic ant on
the periodic simple cubic lattice, as shown in Fig.\ \ref{fig6}.
The data for the blind ant on the periodic lattice were collected on
$30^3$ grids and the average was taken over 449 clusters.
The data for the nonperiodic blind ant were collected for
5000 site clusters and the average was taken over 500 clusters.
The data for the myopic ant were for a $30^3$ lattice, the average being
taken over 500 clusters. The dashed lines shown in the figure are the
least square fits of the data points and all the fits had correlation a
coefficient of greater than 0.999. The least square fitted lines for the
blind ant on the periodic and nonperiodic lattices have the same slope
of $-0.35\pm0.01$ within the statistical error, showing that the boundary
conditions do not affect the value of $d_{s}$ in $d=3$ either. Here, however,
$n(\lambda )$ normalized by the average cluster size for the periodic case
does not seem to have the same amplitude as the nonperiodic case.
The myopic ant on the simple cubic periodic lattice gave a slightly
larger slope of $-0.37\pm0.01$, similarly to the case of $d=2$.
The details of the extracted exponent estimates along with
the values of the exponent obtained from a standard reference \cite{havlin}
are given in Table\ \ref{table1}.  When these results are compared,
we see that they are consistent with each other within the respective errors.

In Fig.\ \ref{fig7}, a log-log plot of $n(\lambda)$ against $\ln\lambda$
for $p>p_{c}$ is shown for both $d=2$ and $d=3$.  In this case,
the substrate becomes more ordered and the $ n(\lambda) $
is expected to scale as $|\ln\lambda|^{d/2-1}$, where $d$ is the Euclidean
dimension of the substrate. Our data show a spiky structure
({\em not} statistical fluctuations) due to the discrete nature of
the eigenvalues.  In the $d=2$ case, the data were collected for
$p=0.75$ and for $p=0.9$ for the clusters on a 100$\times$100 periodic
square grid and the average was taken over 250 and 200 clusters. In $d=3$,
the scaling of $n(\lambda)$ was tested for $p= 0.5$ and the average was taken
over 100 clusters on a $30^3$ periodic grid. In these figures, lines of
expected slope (zero for $d=2$ and 1/2 for $d=3$) are drawn to guide the eye.
As is evident from Fig.\ \ref{fig7}, the data for $d=3$ have a larger slope
than for $d=2$ as expected.

\subsection{Asymptotic behavior of $ \pi(\lambda) $}
As discussed earlier, the function $\pi(\lambda)$ at $p_c$ is expected to
follow the power law $ \pi(\lambda) \sim |\ln\lambda|^{1-2/d_{w}}$,
where $ d_{w} $ is the walk dimension. This exponent is
thus extracted from the slope of the log-log plot of $\pi(\lambda)$
versus $|\ln\lambda|$ as shown in Fig.\ \ref{fig8}. Plotted in this
figure are the results from the blind ant for nonperiodic clusters on
the square and simple cubic lattices and those from the myopic ant for
periodic clusters on the same two lattices (for $\lambda$ near 1 and
{\em not} -1, as discussed in Sec.\ \ref{sec:spectrum}).  The dashed lines
in the figure are the least squares fit and the computed correlation
coefficient was greater than $0.999$ in every case except for the myopic
ant in three dimensions where it was $0.997$. Some data points
toward both ends of $|\ln\lambda|$ were not included where they
correspond to partially filled bins (i.e., only some cluster
realizations have eigenvalues falling in these bins).
The nonperiodic clusters were of size 5000 and, in the periodic case,
grids of size 100$\times$100 in $d=2$ and of $30^3$ in $d=3$ were used.
The average was taken over the same number of clusters as in the
corresponding case of $n(\lambda )$ except for the myopic ant for the
periodic simple cubic lattice where only 200 clusters were averaged.

The estimated value of $1-2/d_w$ for $d=2$ is $0.30 \pm 0.01$ for the
myopic ant on the periodic square grid and $0.27 \pm 0.01$ in the case of
the blind ant on the nonperiodic square grid. (The estimates for the periodic
lattice are actually identical for both the blind and myopic ants.)
In $d=3$ the slope of the least squares fit was found to be $0.46 \pm 0.02$
for the myopic ant on the nonperiodic simple cubic grid and $0.43 \pm 0.01$
for the blind ant for the nonperiodic case. The exponents calculated from
the slopes for the data for the periodic lattice agree very well with the
previously available ones from Ref.\cite{havlin} within the respective
errors.  Although our slope estimates themselves are substantially different
between the periodic and non-periodic clusters, the corresponding exponent
$d_w$ calculated from these slopes do not differ quite as much.
While the remaining deviations are still larger than the statistical errors
and thus a source of concern,
we believe that the periodic lattice results are inherently more trustworthy
and the non-periodic lattice results are somehow reflecting the more severe
boundary effects and/or the effects of the cluster geometry imposed by
arbitrarily stopping the cluster growing algorithm at a predetermined size.
We do not, however, understand why this effect seems to be much more
prominent for $\pi (\lambda )$ than for $n(\lambda )$.

In Fig.\ \ref{fig9} a log-log plot of $\pi(\lambda)$ against $|\ln\lambda|$
for $p>p_{c}$ is presented.  For $ p>p_{c} $,  the function $\pi(\lambda)$
is expected to behave like $|\ln\lambda|^{d/2}$ as $\lambda \rightarrow 1$.
Again in this case we find some structure in the data, and lines of slope 1
and 1.5 are drawn in the figure to guide the eye. For $d=2$, the data were
collected at $p=0.75$ and $p=0.9$, and for $d=3$, the data were collected
for $p=0.5$. The number of clusters over which the average was taken is
the same as the corresponding data for $n(\lambda )$.  Again it is evident
from the figure that the slope of $\pi (\lambda )$ in the case of $d=3$ is
larger than in $d=2$ as expected. The summary of all these results are given
in Table\ \ref{table2}.

\subsection{Subdominant eigenvalue $\lambda_{max}$}
The values of the maximum eigenvalue below one, $\lambda_{max}$, in two
and three dimensions are plotted in Fig.\ \ref{fig10} against the cluster
size.  They were calculated for
clusters of size 100, 200, 400, 800, 1000, and 5000 for the myopic ant
on the triangular lattice and for the blind ant on the square lattice
for cluster sizes of 100, 200, 400, 800, and 5000.  In three dimensions,
$\lambda_{max}$ was found for the blind ant on the simple cubic lattice
of cluster sizes of 100, 400, 1000, and 5000.  The averages for the
cluster sizes 100, 200, 400, 800, 1000, and 5000 were taken over
10000, 5000, 5000, 2500, 2500, and 1000 clusters, respectively.
The accuracy in the numerical evaluation
was better than one part in $10^{6}$ in all cases.

As discussed above the subdominant eigenvalue $\ln \lambda$ scales with
the size of the cluster, $S$, as $|\ln\lambda_{max}| \sim S^{-d_w/d_f}$
at $p=p_c$.  We have extracted this exponent $d_w/d_f$ from the slope
of the log-log plot of $|\ln\lambda_{max}|$ versus $S$ as shown in
Fig.\ \ref{fig10}.
The straight lines drawn are linear least square fits and all of them have
a correlation coefficient of greater than $0.9999$. The statistical errors
are smaller than the symbol sizes in all cases.

In $d=2$ we have obtained a slope magnitude of $1.546 \pm 0.019$ for the
myopic ant on the triangular lattice and a slope of $1.529 \pm 0.004$ for
the blind ant on the square lattice. In comparison, using the value of
$d_s =1.30 \pm 0.002$ as given in Ref.\cite{havlin}, and assuming the
scaling relation $d_s=2d_f /d_w$, we would get the corresponding value of
$d_w/d_f = 1.538 \pm 0.024$.  Thus we see that our estimate computed from
the finite size scaling of the subdominant eigenvalues is in excellent
agreement with the direct simulation result (and the scaling relation used).

In $d=3$ the value of the slope magnitude obtained is $1.508 \pm 0.007$.
Again using the value of $d_s =1.328 \pm 0.006$ from Ref.\cite{havlin} and
the relation $d_s=2d_f/d_w$, the simulation estimate of this quantity would
come out to be $1.506 \pm 0.007$, again in excellent agreement with our
analysis. The summary of these results are given in Table\ \ref{table3}.

In order to check the finite size scaling relation for $|\ln\lambda_{max}|$
for $p$ close to but greater than $p_{c}$, we plot in Fig.\ \ref{fig11},
$|\ln\lambda_{max}| S^{2/d_{s}}$ against $(p-p_{c}) S^\sigma$
for $p=0.64$, 0.65, 0.66 in $d=2$ and for $p=0.33$, 0.34, 0.35 in $d=3$,
where $\sigma =1/(d_f \nu)$.  All the data in $d=2$ are for the blind ant
on square, nonperiodic clusters, and in $d=3$, for simple cubic, nonperiodic
clusters. In $d=2$, the average was taken over 10000, 10000, 5000
and 1000 clusters for cluster sizes of 100, 400, 1000 and 5000, respectively.
In $d=3$, the average was taken over 100, 250, 300 and 1000 clusters for
the cluster sizes of 100, 400, 1000 and 5000, respectively. The statistical
errors are typically of the order $10^{-6}$ and thus less than the symbol
size in each case.

As is evident from the figures, both in $d=2$ and 3, the data for different
values of $p$ collapse onto a single curve, which is the
universal scaling function $f(x)$ (where $x=(p-p_c )S^{\sigma}$),
supporting the scaling form given by Eq.\ (\ref{eq:general}).
The scaling function is numerically fitted to a
quadratic curve of the form $ ax^{2} + bx + c $  as shown by the dashed
lines in the figure. The values for the coefficients $a$,
$b$, $c$ have been numerically estimated as 2.646, 0.893, 2.358 from the
square lattice data and 0.894, 0.276, and 1.044 from those on the simple
cubic lattice, respectively. Even though $f(x)$ is expected to depend only
on the dimensionality, in general there will be lattice-dependent
{\em metrical} factors which will affect the values of the coefficients.

The nonzero value of $c$ indicates that $f(x)$ indeed tends to a nonzero
constant in the limit $x \rightarrow 0$.  This is consistent with the
scaling relationship for $|\ln\lambda_{max}|$ at $p=p_{c}$. Our scaling
prediction for the value of the exponent $z$ is 1.21 for $d=2$ and 1.81 for
$d=3$, respectively, from Eq.\ (\ref{eq:pho}). The value of $z$ was not
estimated numerically since it is valid only in the asymptotic limit
$x \rightarrow \infty$ thus needing the calculation of $\lambda_{max}$
for cluster sizes larger than available due to computational limitations.

\section{Summary}
\label{sec:summary}
In summary, we used the Arnoldi-Saad algorithm \cite{saad} to approximately
diagonalize the hopping transition matrix and successfully extracted
the dynamical exponents $d_{w}$ and $d_{s}$ at the critical value of $p$
for the percolation cluster. We found that, for $p>p_{c}$, the eigenvalue
spectrum reflects the discrete, local lattice structure more clearly,
and mainly because of this there is a spiky
nonstatistical structure in the density of states.  Despite the numerical
difficulty caused by this, we clearly observe the crossover between the
fracton regime (anomalous diffusion) and the phonon regime (Euclidean
diffusion with a long time tail).

We have also shown that the finite size scaling of the subdominant eigenvalue
$\lambda_{max}$ to be the most computationally efficient method to extract
the critical exponent combination $d_w/d_f$ (which is equal to $2/d_s$ for
percolation). The computational efficiency is due to the fact that only the
largest eigenvalue (below one) has to be evaluated.

The method of eigenspectrum analysis as presented here reduces the study
of diffusion in complex, random media to that of a numerical analysis of
large, random matrices.  Such an analysis can be carried out with very high
precision in a systematic way taking the best advantage of modern computing
equipment (with vector and parallel processing).  In fact, this approach was
recently applied by us to a number of problems including the {\em scalar}
elasticity of {\em loop-enhanced} structures \cite{loops} and a study of
the breakdown of the scaling relation $d_s =2d_f/d_w$ for treelike
structures \cite{trees}.  Further applications contemplated include the
problem of diffusion in an external velocity field, a possible model of
Brownian transport in the presence of macroscopic convection.

\acknowledgments
We appreciate fruitful discussions with J. W. Halley, A. Giacometti,
Hans J. Herrmann, R. Muralidhar, and especially D. J. Jacobs.
This work was supported in part by a grant from ONR and we
are grateful to the generous use of computer time at the Supercomputer
Center of the Naval Oceanographic Office.

\newpage
\appendix
\section*{Expansion coefficients $\lowercase{a}_{\lambda}$
for periodic boundaries}
\label{sec:coeff}
We show in this Appendix how the calculation of the expansion coefficients
$a_{\lambda}$ which appear in the function $\pi(\lambda )$ can be obtained
from the {\em velocity} vectors in the case of the periodic boundary
condition.  From Eqs.\ (\ref{eq:coeff1}) -- (\ref{eq:coeff3}), we have
\begin{equation}
a_{\lambda}=\sum_{s} v_{\lambda}(s) {\bf r}(s) \cdot
\sum_{s_{0}} u_{\lambda}(s_{0})\rho(s_0){\bf r}(s_{0}),
\label{eq:exp}
\end{equation}
where ${\bf u_{\lambda}}$ and ${\bf v_{\lambda}}$ are normalized left
and right eigenvectors of ${\bf W}$ for the eigenvalue $\lambda$, ${\bf \rho}$
is the initial probability distribution, and the dot product refers to the
coordinate space as before.

Replacing the coordinate vectors ${\bf r}(s)$ by the velocity vectors
$[{\bf r}(s)-{\bf r}(s_i)]$ where the sites $s_i$ are
the nearest neighbors of the site $s$, the first of the two factors in
the above equation becomes
\begin{equation}
\sum_{s}v_{\lambda}(s){\bf r}(s) = \sum_{s}v_{\lambda}(s)
\frac{1}{z(s)}\sum_{i=1}^{z(s)}[{\bf r}(s)-{\bf r}(s_{i})]
+ \sum_{s} v_{\lambda}(s)\frac{1}{z(s)}\sum_{i=1}^{z(s)} {\bf r}(s_{i}) .
\end{equation}
Here $z(s)$ is the number of occupied neighbors of the site $s$ in the
case of the myopic ant. For the blind ant, $z(s)$ stands for the lattice
coordination number and, if the $i$th neighbor is not a cluster site, then
$s_i \equiv s$ is understood.

The second term on the right can be simplified by exchanging the summations
over $s$ and $s_i$:
\begin{eqnarray}
 \nonumber
\sum_{s} v_{\lambda}(s)\frac{1}{z(s)}\sum_{i=1}^{z(s)}{\bf r}(s_{i})
 & = & \sum_{s'}{\bf r}(s')\sum_{i=1}^{z(s')}
       \frac{1}{z(s_i ')} v_{\lambda} (s_i ') \\
 \nonumber
 & = & \sum_{s'}{\bf r}(s'){\lambda} v_{\lambda}(s') \\
 & = & {\lambda}\sum_{s'}{\bf r}(s') v_{\lambda}(s') ,
\end{eqnarray}
where we used the relation
\begin{equation}
\sum_{i=1}^{z(s')}\frac{1}{z(s_i ')} v_{\lambda}(s_i ') =
\sum_{s"} W(s',s") v_{\lambda}(s") .
\end{equation}
Therefore we get
\begin{equation}
\sum_{s} v_{\lambda}(s){\bf r}(s) = \frac{1}{(1-{\lambda})}\sum_{s}
v_{\lambda}(s)\frac{1}{z(s)}\sum_{i=1}^{z(s)}[{\bf r}(s)-{\bf r}(s_{i})].
\end{equation}
Similarly the second factor in Eq.\ (\ref{eq:exp}) is
\begin{equation}
\sum_{s} u_{\lambda}(s){\bf r}(s)\rho(s) = \frac{1}{(1-{\lambda})}
\sum_{s} u_{\lambda}(s)\rho(s)\frac{1}{z(s)}
\sum_{i=1}^{z(s)}[{\bf r(s)}-{\bf r}(s_{i})].
\end{equation}
Once in this form, the absolute coordinates are not required; rather,
only the nearest neighbor {\em velocities} are.  In the calculation of
the expansion coefficients $a_{\lambda}$, then, we only need to keep
track of the velocity vectors in the case of the periodic boundary
conditions.

\newpage
\begin{figure}
\caption{Density of eigenvalues $n(\lambda )$ for the myopic ant on
the square lattice cluster of size 200 at $p=0.593\approx p_{c}$ and $p=0.8$.
This figure shows the build up of $n(\lambda )$ near $\lambda =\pm1$
at $p=0.593$ and the flattening of $n(\lambda )$ for $p>p_c$.
\label{fig1}}
\end{figure}

\begin{figure}
\caption{Density of eigenvalues $n(\lambda )$ at $p=0.593$ for the blind ant
on small square lattice clusters of size 800 and 100 sites. This figure
shows the peaks due to the local structures of the cluster as well as
the increase in $n(\lambda )$ near one for larger size clusters.
\label{fig2}}
\end{figure}

\begin{figure}
\caption{Illustration of the eigenvectors which contribute to the peaks
in the eigenvalue spectrum at $p=0.75$ and $p=0.5$ using a 100-site cluster
on the square lattice (sites shown with dots).
Some (not all) such eigenvectors are chosen and the sites of
nonzero amplitudes for eigenvectors with eigenvalue $\lambda =0.75$
are marked by the symbol $+$ and those of $\lambda =0.5$ are marked
by $\bigtriangleup$.
\label{fig3}}
\end{figure}

\begin{figure}
\caption{Density of eigenvalues $n(\lambda )$ normalized by the cluster size
for the blind ant on nonperiodic clusters on square and simple cubic lattices.
The upper curve corresponds to $d=3$ and the lower one to $d=2$. The symbols
$+$, $\bigtriangleup$ and $\circ$ correspond to the cluster sizes 400,
800 and 5000, respectively.
\label{fig4}}
\end{figure}

\begin{figure}
\caption{Log-log plot of $n(\lambda)$ against $|\ln\lambda|$ in $d=2$
at $p=0.593\approx p_{c}$. The symbols $\bigtriangleup$ and $+$ correspond
to the blind ant on periodic and nonperiodic square lattices and the symbol
$\circ$ corresponds to the myopic ant on periodic square lattice.
\label{fig5}}
\end{figure}

\begin{figure}
\caption{Log-log plot of $n(\lambda)$ against $|\ln\lambda|$ in $d=3$
at $p=0.312\approx p_{c}$. The symbols $\bigtriangleup$ and $+$ correspond
to the blind ant on periodic and nonperiodic simple cubic lattices and
the symbol $\circ$ corresponds to the myopic ant on periodic simple cubic
lattice.
\label{fig6}}
\end{figure}

\begin{figure}
\caption{Log-log plot of $n(\lambda)$ against $|\ln\lambda|$ in $d=2$ and 3
for $p>p_{c}$.  The symbols $\circ$ and $\bigtriangleup$ stand for the
blind ant on the periodic square lattice at $p=0.75$ and $p=0.9$,
respectively, and the symbol $+$ stands for the blind ant on the periodic
simple cubic lattice at $p=0.5$. The lines of slopes zero and $1/2$ are
drawn to guide the eye.
\label{fig7}}
\end{figure}

\begin{figure}
\caption{Log-log plot of $\pi(\lambda)$ against $|\ln\lambda|$ in $d=2$ and 3
at $p=p_{c}$.  The symbols $\bigtriangleup$ and $\circ$ stands for the
myopic ant on the periodic square lattice and the blind ant on the
nonperiodic square lattice, respectively, and the symbols $\times$ and
$+$ correspond to the myopic ant on the periodic simple cubic lattice
and the blind ant on the nonperiodic simple cubic lattice, respectively.
\label{fig8}}
\end{figure}

\begin{figure}
\caption{Log-log plot of $\pi(\lambda)$ against $|\ln\lambda|$ in $d=2$ and 3
for $p>p_{c}$.  The symbols $\circ$ and $\bigtriangleup$ stand for the
blind ant on the periodic square lattice at $p=0.75$ and $p=0.9$, respectively.
The symbol $+$ stands for the blind ant on the periodic simple cubic lattice
at $p=0.5$.
\label{fig9}}
\end{figure}

\begin{figure}
\caption{Log-log plot of $|\ln\lambda_{max}|$ against the size $S$
of the cluster. The symbol $\circ$ corresponds to the myopic ant on the
triangular lattice.  The symbols $\bigtriangleup$ and $+$ correspond to the
blind ant on the square and simple cubic lattices, respectively.
\label{fig10}}
\end{figure}

\begin{figure}
\caption{Finite size scaling of $|\ln\lambda_{max}|$ with respect to
the size of the cluster and $(p-p_{c})$. Quadratic fits for the universal
scaling functions $f(x)$ are shown in dashed lines, where
$x=(p-p_c)S^{\sigma}$ and the upper curve corresponds to $d=2$, the lower
to $d=3$.
\label{fig11}}
\end{figure}

\newpage
\begin{table}
\caption{Estimates of $d_{s}/2-1$ from the density of eigenvalues $n(\lambda
)$}
\label{table1}
\begin{tabular}{cccc}
d & This work & Using $d_s$ from \protect\cite{havlin}& $d/2-1$ (for $p>p_c$)\\
\hline
  2 & $-0.35\pm 0.01$& $-0.347\pm 0.013$ & $ 0 $ \\
  3 & $-0.35\pm 0.01$& $-0.336\pm 0.003$ & $ 1/2 $
\end{tabular}
\end{table}

\begin{table}
\caption{Estimates of $1-2/d_{w}$ from $\pi(\lambda)$}
\label{table2}
\begin{tabular}{cccc}
  d & This work & Using $d_w$ from \protect\cite{havlin}& $d/2$ (for $p>p_c$)\\
\hline
  2 & $0.30\pm 0.01$& $0.30\pm 0.01$ & $ 1 $ \\
  3 & $0.46\pm 0.01$& $0.46\pm 0.03$ & $ 3/2 $
\end{tabular}
\end{table}

\begin{table}
\caption{Estimates of $d_w /d_f$ from $\ln\lambda_{max}$}
\label{table3}
\begin{tabular}{ccccc}
d & Ant type & This work & Using $2/d_s$ from \protect\cite{havlin} \\
\hline
2 & myopic & $1.546 \pm 0.019$ & $1.538 \pm 0.024$  & \\
2 & blind  & $1.529 \pm 0.004$ & $1.538 \pm 0.024$  &\\
3 & blind  & $1.508 \pm 0.007$ & $1.506 \pm 0.007$  &
\end{tabular}
\end{table}

\newpage
\begin{references}
\bibitem{random}{\em Random walks and Their Applications in the Physical
and Biological Sciences}, edited by Michael F. Shlesinger and Bruce J.West,
AIP Conf. Proc. No. 109 (AIP, New York, 1984).
\bibitem{straley}J. P. Straley in {\em Electrical and Optical Properties
of Inhomogeneous Media }, edited by J.C. Garland and D.B. Tanner, AIP
Conf. Proc. No. 40 (AIP, New York, 1990), and references therein.
\bibitem{degennes}P.G. de Gennes, {\em Scaling Concepts in Polymer Physics}
(Cornell University, Ithaca, NY, 1979), and references therein.
\bibitem{des}J. des Cloiseaux and G. Jannik, {\em Polymers in solution:
Their Modelling and Structure} (Oxford, New York, 1990), and references
therein.
\bibitem{stauffer}D. Stauffer and A. Aharony, {\sl Introduction to
Percolation Theory} (Taylor and Francis, London, 1992).
\bibitem{mandelbrot}B. Mandelbrot, {\em Fractal Geometry of Nature}
(Freeman, San Francisco, 1982).
\bibitem{havlin}S. Havlin and D.Ben-Avraham, Adv. Phys. {\bf 36},
695 (1987), and references therein.
\bibitem{vankampen}N. G. van Kampen, {\em Stochastic Processes
in Physics and Chemistry} (North-Holland, Amsterdam, 1981).
\bibitem{gefen}Y. Gefen, A. Aharony, and S. Alexander, Phys. Rev. Lett.
{\bf 50}, 77 (1983).
\bibitem{alder}B. J. Alder and W. E. Alley, J. Stat. Phys.
{\bf 19}, 341 (1978); Physica A {\bf 121}, 523 (1983).
\bibitem{ernst}M. H. Ernst and A. Weyland, Phys. Lett. {\bf 34A}, 39 (1971).
\bibitem{nakanishi}H. Nakanishi, S. Mukherjee and N. H. Fuchs, Phys. Rev. E
{\bf 47}, R1463 (1993).
\bibitem{fuchs}N. H. Fuchs and H. Nakanishi, Phys. Rev. A
{\bf 43}, 1721(1991).
\bibitem{jacobs}D. J. Jacobs and H. Nakanishi, Phys. Rev. A
{\bf 41}, 706 (1990).
\bibitem{alexander}S. Alexander and R. Orbach, J. Phys. Lett. (Paris)
{\bf 43}, 625 (1982).
\bibitem{saad}Y. Saad, Linear Algebra Appl. {\bf 34}, 269 (1980);
see also W.E. Arnoldi, Quart. Appl. Math. {\bf 9}, 17 (1951).
\bibitem{nakayama}K. Yakubo and T. Nakayama, Phys. Rev. B {\bf 40}, 517 (1989).
\bibitem{loops}H. Nakanishi, Physica A {\bf 196}, 33 (1993).
\bibitem{trees}H. Nakanishi and H. J. Herrmann, J. Phys. A
{\bf 26}, 4513 (1993).
\end{references}
\end{document}